\documentclass{emulateapj}
\usepackage{color, hyperref, epsfig}
\usepackage{apjfonts, natbib}

\newcommand{\swift}{\emph{Swift}}
\newcommand{\colnh}{\ensuremath{N_\mathrm{H}}}
\newcommand{\ha}{H\ensuremath{\alpha}}

\newcommand{\feii}{Fe\,{\footnotesize II}}

\newcommand{\lum}{erg~s\ensuremath{^{-1}}}
\newcommand{\lsun}{\ensuremath{L_{\odot}}}
\newcommand{\lbol}{\ensuremath{L\mathrm{_{bol}}}}
\newcommand{\lpeak}{\ensuremath{L\mathrm{_{peak}}}}

\newcommand{\lw}{\ensuremath{L\mathrm{_{W2}}}}
\newcommand{\lha}{L\mathrm{_{H\ensuremath{\alpha}}}}
\newcommand{\msun}{\ensuremath{M_{\odot}}}
\newcommand{\kms}{\ensuremath{\mathrm{km~s^{-1}}}}
\newcommand{\mbh}{\ensuremath{M_\mathrm{BH}}}
\newcommand{\wise}{\emph{WISE}}
\newcommand{\spitzer}{\emph{Spitzer}}
\newcommand{\neowise}{\emph{NEOWISE}}
\newcommand{\tsub}{\ensuremath{T_\mathrm{sub}}}
\newcommand{\tdust}{\ensuremath{T_\mathrm{dust}}}
\newcommand{\rsub}{\ensuremath{R_\mathrm{sub}}}
\newcommand{\erad}{\ensuremath{E_\mathrm{rad}}}
\newcommand{\edust}{\ensuremath{E_\mathrm{dust}}}
\newcommand{\fdust}{\ensuremath{f_\mathrm{dust}}}
\begin{document}

\title{
Infrared Echo and late-stage rebrightening of nuclear transient PS1-10adi:
Exploring Torus by tidal disruption event in active galactic nuclei
}
\author{
Ning~Jiang\altaffilmark{1,2},
Tinggui~Wang\altaffilmark{1,2},
Guobin~Mou\altaffilmark{3},
Hui~Liu\altaffilmark{1,2}, 
Liming~Dou\altaffilmark{4},
Zhenfeng~Sheng\altaffilmark{1,2},
Yibo~Wang\altaffilmark{1,2}
}
\altaffiltext{1}{Key laboratory for Research in Galaxies and Cosmology,
Department of Astronomy, University of Science and Technology of China,
Chinese Academy of Sciences, Hefei, Anhui 230026, China; ~jnac@ustc.edu.cn}
\altaffiltext{2}{School of Astronomy and Space Sciences,
University of Science and Technology of China, Hefei, Anhui 230026, China}
\altaffiltext{3}{School of Physics and Technology, Wuhan University, Wuhan 430072, China}
\altaffiltext{4}{Center for Astrophysics, Guangzhou University, Guangzhou 510006, China}

\begin{abstract}
Tidal disruption events (TDEs) in active galactic nuclei (AGNs) have been overlooked 
for a long time but tentatively been investigated recently.
We report the discovery of a long-lasting luminous mid-infrared (mid-IR) flare in PS1-10adi,
which is a newly-identified highly energetic transient event occurred in AGN.
The IR luminosity of PS1-10adi, as well as other analogous events, are 
at least one order of magnitude higher than all known supernova, 
but can be well interpreted as the dust echoes of TDEs, 
whose ultra-high IR energy is reprocessed from the dusty torus around the black hole.
The torus dust is sublimating during the early stage of the outburst and probably lead to
the observed rapid emergence of \feii\ lines.
Moreover, the UV-optical rebrightening and contemporaneous X-ray onset after $\sim1500$ 
rest-frame days since the optical peak is also an intriguing feature of PS1-10adi, 
which could be attributed to the interaction between the high-velocity outflow and torus.
We suggest that the luminous IR echo is a very typical phenomenon of TDEs in AGNs
and may provide us an ideal opportunity to explore the torus properties.

\end{abstract}


\section{Introduction}

The stellar or gas kinematics on nearby galaxies in the past two decades 
have established that supermassive black holes (SMBHs), with masses of $10^{6-10}$~\msun, 
are universal in the centers of  galaxies with massive bulges.
Furthermore, the tight correlations between the BH mass (\mbh) and various bulge properties
indicate an attractive co-evolutionary growth of SMBHs and their host galaxies
(see reviews by \citealt{KH13} and \citealt{HeckBest2014}).
The widespread picture is not unquestionable but still competed
by some alternative views which claim that their
relation can emerge as the result of a statistical convergence
process without a physical coupling (e.g., \citealt{Peng2007};
\citealt{Jahnke2011}) or the dark matter might be influential
(e.g., \citealt{Ostriker2000}; \citealt{Zhao2002}; \citealt{Bogdan2015}).
Whatever, SMBHs are normally believed to accumulate their mass through the phase of active 
galactic nuclei (AGN), during which the BHs are efficiently accreting surrounding materials.
AGNs themselves are also the best evidence of the existence of SMBHs in distant galaxies,
which are otherwise beyond the current capabilities for direct dynamical measurement.

Nevertheless, the detection of SMBHs in normal galaxies are still very 
difficult provided that vast majority of them are quiescent and faraway. 
Although some new approaches seem promising to detect SMBHs at cosmological distances,
such as gravitational lensing (e.g., \citealt{Mao2001};
\citealt{Hezaveh2015}; \citealt{Chen2018}) and future gravitational
wave observations of SMBH mergers or the inspiral
of compact stellar remnants consumed by SMBH, they are
not so efficient and practical for the time being.
Heaven will always leave a door open, the so-called tidal disruption event (TDE)
could offer us the greatest chance to catch sight of these dormant SMBHs.
The TDE happens if a star in the galaxy wanders too close to the central black hole, 
the star can be ripped apart when the tidal force exceeds its self-gravity
and roughly half mass of the star will be accreted on to the black hole while the 
remaining half may be ejected by the simple classical theory (Rees 1988).
Recent works reveal that the real accreting fraction could be much lower since most of
the mass may join in an outflow (e.g., \citealt{Metzger2016}).
A luminous flare of electromagnetic radiation is expected during this process
with emission peaks in the UV or soft X-rays and a characteristic $t^{-5/3}$
decline on timescale of months to years (\citealt{Rees1988}; \citealt{Phinney1989}).
The TDE event rate is estimated to be 
$10^{-4}$ to $10^{-5}\rm galaxy^{-1} yr^{-1}$ and might be highest 
in nucleated dwarf galaxies (\citealt{WM2004}, \citealt{Stone2016}).

Observationally, TDEs are though first discovered in X-ray bands serendipitously,
the field has not experienced explosive growth until the past decade, 
benefited from a batch of dedicated optical time-domain surveys 
(see \citealt{Komossa15} as a review).
To avoid any potential contaminations stemmed from variability of AGN,
traditional searching of TDEs care about solely normal galaxies and 
neglected active galaxies (e.g., \citealt{vanv2011}).
However, TDEs might also occur in AGNs and the rates are expected to be even 
high (\citealt{Karas2007}), not to mention that TDEs ware actually first 
proposed out as a possible power source of AGNs and quasars (\citealt{Hills1975}).
Recently, \citealt{Blanchard2017} have claimed a TDE candidate PS16dtm found in 
the Panoramic Survey Telescope and Rapid Response System (Pan-STARRS).
This event occurs in a Seyfert~1 galaxy with \mbh$\sim10^6$~\msun, 
and display some distinctive features, such as a plateau phase, disappeared 
prior X-ray emission and newly-emerging strong \feii\ emission.
Soon thereafter, Kankare et al. (2017, hereafter K17) have reported 
the other high energetic Pan-STARRS transient event named PS1-10adi with a
total radiated energy of $\sim2.3\times10^{52}$~erg.
The rapid smooth brightening light curve and persistently narrow spectral lines 
over $\sim$3 yr of PS1-10adi is inconsistent with known types of recurring 
BH variability, yet may be linked to a TDE or supernova (SNe) in the nuclear region
powered by shock interaction between expanding material and large quantities 
of surrounding dense matter.
K17 have even unveiled possibly a population of such kind of events 
in the centers of active galaxies yet to be further explored.

Encouraged by our successful detection of a mid-infrared (mid-IR) flare in PS16dtm and
valuable implications learned from it (\citealt{Jiang17}), 
we have also checked the mid-IR variability of PS1-10adi and discovered a long-lasting 
flare not surprisingly in the location coincident with PS1-10adi using data from the 
$\emph{Wide-field Infrared Survey Explorer}$ (\wise; \citealt{Wright2010}; 
\citealt{Mainzer2014}).
In addition, we have noticed an odd late-stage rebrightening in the UV-optical
light curve as well as a corresponding mid-IR signal.
The mid-IR emission in radio-quiet AGNs is generally thought to be associated with 
thermal emission from dust at parsec-scale distances from the central SMBH. 
On the other hand, an optically thick obscuring medium in the equatorial plane 
(commonly dubbed the "torus") is needed in the unified model, 
the current paradigm suggests that the IR-emitting dust is the same medium as 
the obscuring material. 
This is reflected by the widespread use of torus models in order to reproduce IR 
spectral energy distributions (SEDs, see a recent review on torus in \citealt{Netzer2015}.
Basing on the fact of a pre-existing dusty torus, mid-IR echoes are predictable 
for those TDEs in active galaxies, which would expose the torus itself dynamically.
The ubiquitous luminous IR echoes associated with TDEs in AGNs
may be one of the most notable characteristics distinguished from SNe
(see \S\ref{energy}).

In this paper, we will try to analyze all of these new recognized phenomenons
in PS1-10adi and understand them under the context of TDEs in AGNs.
We assume a cosmology with $H_{0} =70$ km~s$^{-1}$~Mpc$^{-1}$, $\Omega_{m} = 
0.3$, and $\Omega_{\Lambda} = 0.7$.  At a redshift of $z=0.203$, PS1-10adi
has a luminosity distance of 996.5~Mpc.

\section{Mid-IR Light Curves}


The \wise\ has performed a full-sky imaging survey in four broad mid-IR
bandpass filters centered at 3.4, 4.6, 12 and 22~$\mu$m (labeled W1-W4)
from 2010 February to August.
The solid hydrogen cryogen used to cool the W3 and W4 instrumentation
was depleted later and it was placed in hibernation in 2011 February.
\wise\ was reactivated and renamed \neowise-R since 2013 October, using only W1 and W2,
to hunt for asteroids that could pose as impact hazard to the Earth.
\wise\ has scanned a specific sky area every half year and thus yielded 10-11 times
of observations for every object up to now.
For each epoch, there are 10-20 single exposures.  
We use only the best quality single frame images by selecting only detections
with data quality flag 'qual\_frame'>0.

\begin{figure}
\centering
\includegraphics[width=9cm]{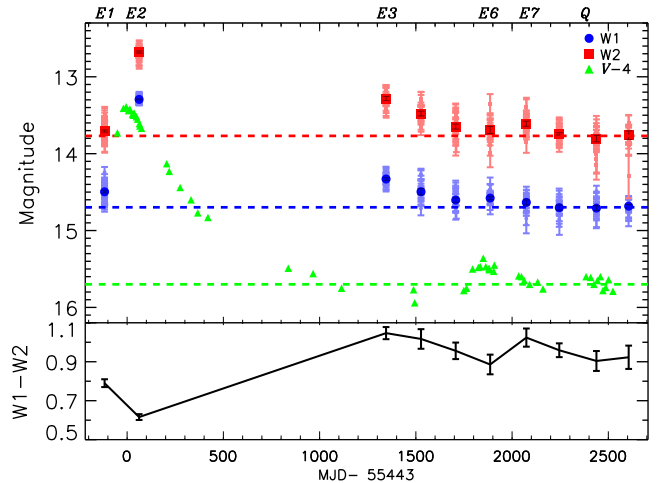}
\caption{
The \wise\ W1 ($3.4\mu$m, blue squares) and W2 ($4.6\mu$m, red triangles) light curves 
of PS1-10adi.
The horizontal dashed lines represent the background magnitudes estimated
from the latest three epochs. 
The $V$-band light curve is also plotted for comparison (green triangles).
The W1-W2 color is shown in the bottom panel.
}
\label{wiselc}
\end{figure}

PS1-10adi is a point-like source in \wise\ images without any
potential contamination within 10\arcsec. 
We have collected all of its public \wise\ photometric data, including from ALLWISE and
\neowise-R catalogs. The data are distributed in 10 epochs at intervals of about 
six months from 2010 May 9 (MJD=55,325) up to 2017 Oct 22 (MJD=58,048). 
The light curves in W1 and W2 are presented in Figure~\ref{wiselc}.
For ease of comparison, its optical light curve ($V$-band) has also been overplotted.
First, no reliable variability is detected within each epoch and thus
we then average these data to obtain a mean value at each epoch following our
previous works (Jiang et al. 2012; 2016). 
We noticed that the first \wise\ epoch ("E1") is 98 days earlier and 
the second epoch ("E2") is 82 days after the optical discovery date,
or -117~(rest-frame -97) days and 63~(52) days since the optical peak.
PS1-10adi has began brightened by $\sim1$~mag in W1 and W2 at E2 relative to E1.
Unfortunately the \wise\ went into hibernation soon and hasn't yielded the third 
observation until 2014 May, resulting in a gap of $1282$ days between E2 and E3.
The light curve presents a declining trend at E3 and thereafter and keeps at 
a constant level in the near past.
In order to isolate the host galaxy and AGN emission (or background emission as a
whole) from the outburst phase, We take the last three epochs as the quiescent state, 
which contributes an averaged magnitude of $14.70\pm0.02$ and $13.77\pm0.03$ 
at W1 and W2, respectively.
The original and background-subtracted magnitudes are shown in Table~\ref{tbl:data}.
It's interesting to note that PS1-10adi is already brighter at E1, 
by $0.20\pm0.02$ and $0.06\pm0.03$ magnitudes in W1 and W2, respectively. 
The W1-W2 color has generally been used as a rough assessment of the AGN   
or stellar dominance of a galaxy, for instance, W1-W2>0.8 for AGN-like
or W1-W2<0.5 for galaxy-like (\citealt{Stern2012}; \citealt{Yan2013}).
The W1-W2 at E1 is $0.79\pm0.02$, significantly lower than $0.93\pm0.05$ at 
the last three epochs (see bottom panel of Figure~\ref{wiselc}), 
indicating likely that it is experiencing an extra high energy input in addition to AGN at E1.
Therefore, we think that the flux excess at E1 is not simply due to AGN variability.

K17 has presented UV-optical-IR light curves of PS1-10adi
up to $\sim2200$ days since the peak (see their Supplementary Figure 1).
When we look through them carefully, it's worthwhile to find that
there's a tiny yet notable hump at MJD$\approx57,293$. For ease of convenience,
we call the first major flare as the "primary outburst" and the second peak
as "rebrightening".
Interestingly, the rebrightening looks also occurs subsequently in W1 and W2, 
in which the W2 lags behind W1 by about half year.
The W1 and W2 magnitudes have brightened by $0.12\pm0.03$~$(4\sigma$) 
and $0.08\pm0.05$~$(1.6\sigma)$ at E6, while by $0.06\pm0.04$~$(1.5\sigma)$ and 
$0.16\pm0.04$~$(4\sigma)$ at E7, respectively.

\begin{deluxetable*}{ccccccccccc}
\tabletypesize{\scriptsize}
\setlength{\tabcolsep}{0.02in}
\tablecaption{{\it {\it WISE}}\ Data
\label{tbl:data}}
\tablewidth{0pt}
\tablehead{
\colhead{Epoch} & \colhead{MJD} & \colhead{W1m} & \colhead{W2m} & \colhead{W1} & \colhead{W2} & \colhead{W1-W2} &  \colhead{log$L_{\rm W1}$} & \colhead{log$L_{\rm W2}$} & \colhead{\tdust} & \colhead{log$L_{\rm dust}$}  \\
\colhead{(1)} & \colhead{(2)}  & \colhead{(3)}  & \colhead{(4)} &
\colhead{(5)} & \colhead{(6)}  & \colhead{(7)}  & \colhead{(8)} &
\colhead{(9)} & \colhead{(10)} & \colhead{(11)} }
\startdata
E1 & 55325 & $14.50\pm0.01$ & $13.71\pm0.01$ & $16.42\pm0.02$ & $16.83\pm0.03$ & -0.41& 42.9 & 42.4 & ..   & ...   \\
E2 & 55505 & $13.29\pm0.01$ & $12.68\pm0.01$ & $13.64\pm0.01$ & $13.17\pm0.03$ & 0.47 & 44.1 & 43.9 & 1972 & 44.3  \\
E3 & 56788 & $14.33\pm0.02$ & $13.28\pm0.03$ & $15.68\pm0.02$ & $14.39\pm0.04$ & 1.29 & 43.2 & 43.3 & 886  & 43.5  \\
E4 & 56969 & $14.50\pm0.04$ & $13.48\pm0.03$ & $16.42\pm0.04$ & $15.50\pm0.04$ & 1.37 & 42.9 & 43.1 & 843  & 43.2  \\
E5 & 57150 & $14.60\pm0.03$ & $13.65\pm0.03$ & $17.31\pm0.03$ & $16.09\pm0.05$ & 1.22 & 42.6 & 42.7 & 927  & 42.8  \\
E6 & 57329 & $14.58\pm0.02$ & $13.69\pm0.05$ & $17.03\pm0.03$ & $16.60\pm0.06$ & 0.43 & 42.7 & 42.5 & 2114 & 43.0  \\
E7 & 57517 & $14.63\pm0.03$ & $13.61\pm0.03$ & $17.75\pm0.04$ & $15.77\pm0.05$ & 1.98 & 42.4 & 42.8 & 620  & 43.1  \\
Q  & 57881 & $14.70\pm0.02$ & $13.77\pm0.03$ & ...            & ...            & 0.93 & ... & ... & ...  & ...  
\enddata
\tablecomments{
Column~(1): observational epochs.
Column~(2): median modified Julian Date.
Column~(3)-(4): median W1 and W2 magnitudes.
Column~(5)-(6): background-subtracted W1 and W2 magnitudes.
Column~(7): background-subtracted W1-W2 color.
Column~(8)-(9): background-subtracted W1 and W2 monochromatic luminosity.
Column~(10): dust temperature inferred from W1-W2 color assuming blackbody radiation.
Column~(11): the logarithmic integrated dust blackbody luminosity.
}
\end{deluxetable*}

\section{Evaporation and Echo of Torus Dust}
\label{echo}

\subsection{Infrared Echo of Torus Dust}
The IR flares detected in TDEs thus far are all successfully translated into
dust echoes (Jiang et al. 2106, 2017; Dou et al. 2016, 2017; \citealt{vanv2016}).
The dust, if there, in the vicinity of SMBHs will certainly absorb the primary emission 
of TDEs (soft X-ray, UV and optical) and reprocess them into IR, 
so we expect the flare in the IR would lag behind the optical (\citealt{Lu2016}).
However, in some cases, the mid-IR variability could be detectable even earlier 
than the optical band and thus the echo interpretation is not so obvious. 
For example, we recently found a mid-IR flare that started $\sim11$ days before the 
optical detection of a TDE candidate PS16dtm (\citealt{Jiang16}).
Interpreting the mid-IR flare as a dust echo requires close pre-existing dust with 
a high covering factor, which is fully qualified with the geometrically and optically
thick torus in AGNs.
The case in PS1-10adi looks even more extreme, in which 
the mid-IR brightening is detectable 118 (or rest-frame 98) days earlier than the 
optical discovery. Can it be explained by the scenario of PS16dtm as well?

\begin{figure}
\centering
\includegraphics[width=8.5cm]{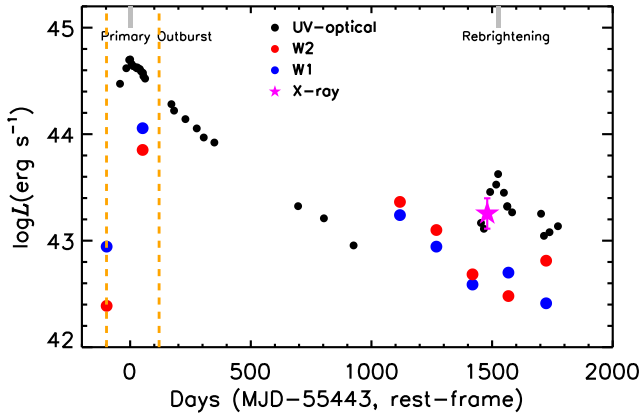}
\caption{
The luminosity evolution of PS1-10adi. 
The black, blue and red dots represent the UV-optical blackbody fitted luminosity,
\wise\ W1 and W2 luminosity respectively. We have marked the peak time of
the primary outburst and rebrightening.  The X-rays detected during the 
rebrightening is plotted as a magenta five-pointed star.
The region between the two orange dashed lines denotes the time
when the torus dust is sublimating. 
}
\label{bol}
\end{figure}
\smallskip

The AGN property before the outburst is unfortunately missing due to lack of observations,
yet can be supposed to be almost the same as the post-flare state.
For the sake of an estimate, we have fitted the optical spectra at +1600 days 
(see Supplementary Figure 2 of K17) and obtained that the broad \ha\ luminosity ($\lha$) 
is $8.8\times10^{41}$~\lum\ with a full width at half maximum (FWHM) of 3,676~\kms. 
By virtue of the empirical relation of \mbh\ in AGNs estimated from the luminosity and 
FWHM of broad \ha\ line (\citealt{GH05}), the \mbh\ is $2.7\times10^7$~\mbh.
Using the bolometric correction given by \citealt{GH07}, the bolometric luminosity (\lbol) 
is $2.34\times10^{44}(\lha/10^{42})^{0.86}=2.1\times10^{44}$~\lum.
The peak background-subtracted \lbol\ associated with the flare inferred from 
the UV-optical emission is $\sim5\times10^{44}$~\lum\ (see Figure~\ref{bol}), 
thus the total peak \lbol\ is $\sim7\times10^{44}$~\lum\ provided that all of the radiation
is originated from the BH accretion.

\begin{figure}
\centering
\includegraphics[width=8.cm]{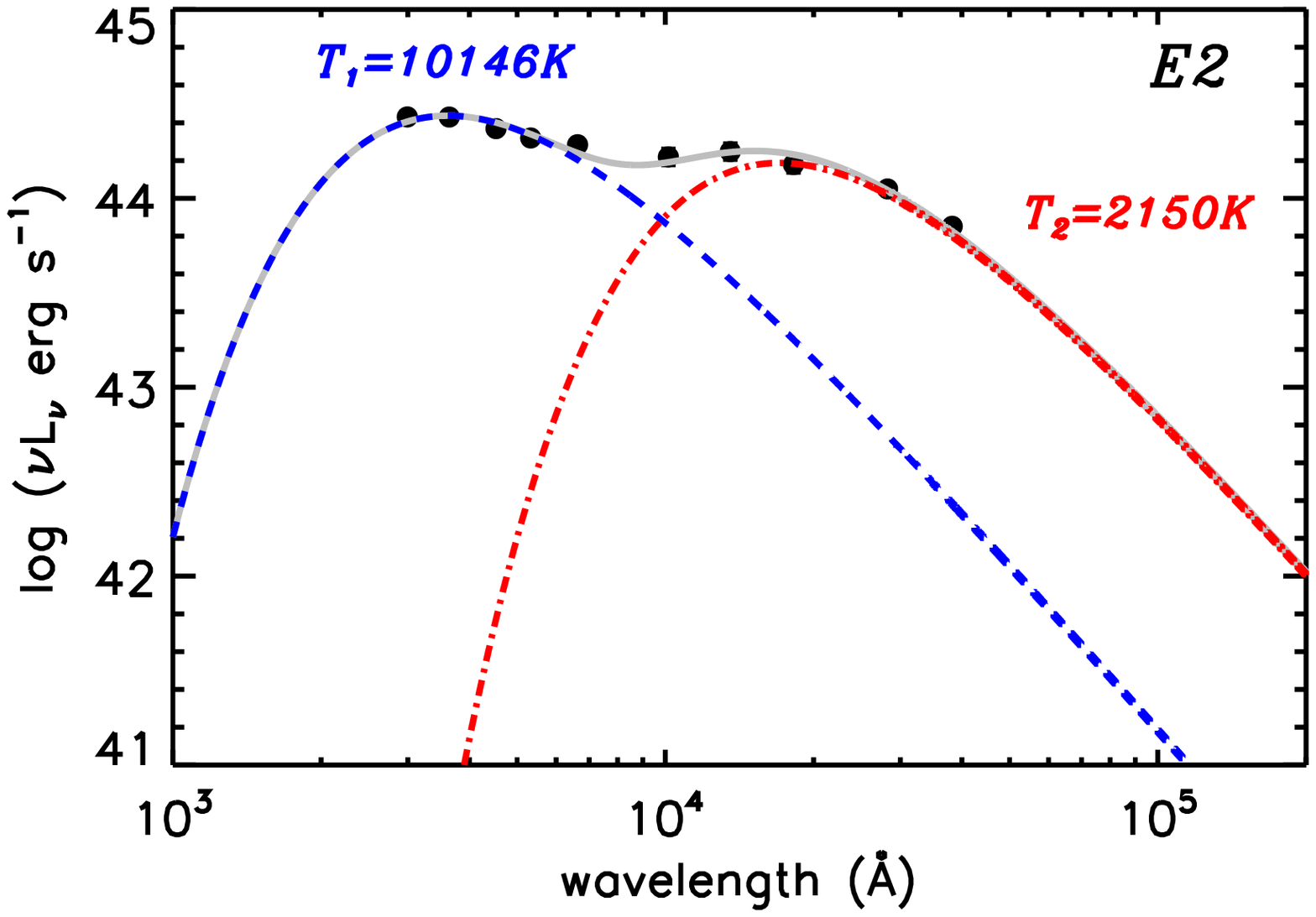}
\includegraphics[width=8.cm]{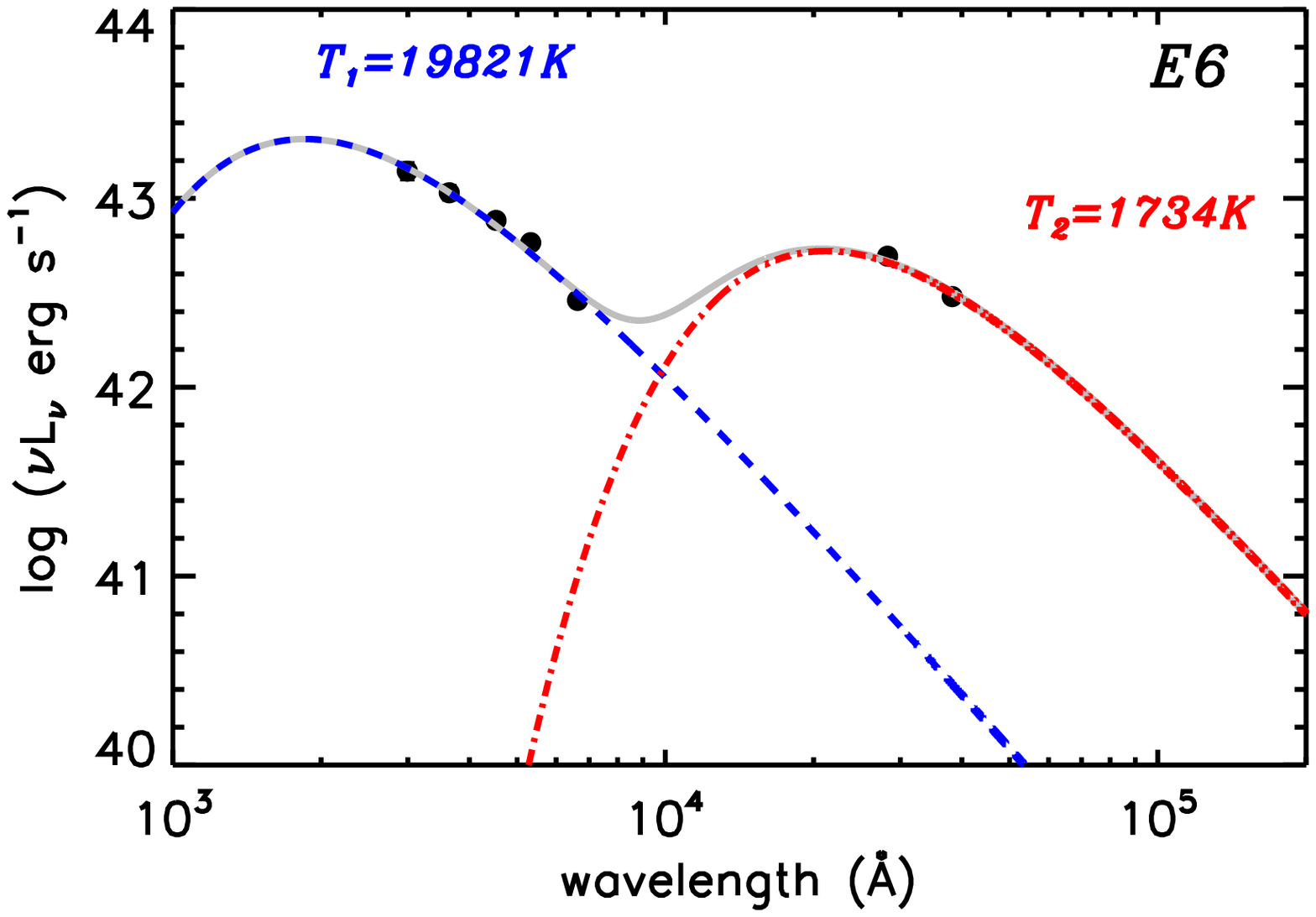}
\caption{
The background-subtracted SED of PS1-10adi at E2 (MJD$\approx$55,505) and E6
(MJD$\approx$57,323).
The optical-NIR photometry is drawn from supplementary table~2 and table~3 of K17.
We have tried to fit the SED using double-blackbody model, one for the  UV-optical
emission (blue dashed line) and the other for the dust emission (dot-dashed line).
}
\label{sed}
\end{figure}

In order to assess if the energy budget for the dust echo is reasonable,
we have fitted the background-subtracted UV-optical-IR SEDs at different epochs
using double-blackbody model, one for the original outburst emission dominating 
UV-optical band and the other for the reprocessed dust emission peaked in the IR
(see an example in the top panel of Figure~\ref{sed}).
We take the integrated luminosity of the blue component as the \lbol\ 
(see its evolution in Figure~\ref{bol}). 
Since there is no reliable detection of residual UV-optical emission at E1, 
a single blackbody is adopted and yields an ultra-high dust temperature of $\sim10^7$~K.
The value seems absolutely unreasonable probably because of 
a non-equilibrium condition, too simplified model and insensitivity to the 
slope of Rayleigh-Jeans tail for very high blackbody temperature. 
Therefore, we believe that the dust at E1 is sublimating in spite of the very uncertain 
dust temperature.

The double-blackbody fitting at E2 yields a dust temperature (\tdust) of $2,150\pm430$~K,
that is consistent with the dust sublimation temperature within error.
Assuming that the equilibrium \tdust\ of a dust grain with radius $a$ at distance of $R$
from the heating source is determined by the
balance between the radiative heating by UV-optical photons
and the thermal re-emission in the IR:
\begin{equation}
  \frac{L_{\rm bol}}{4\pi R^2} \pi a^2  = 4\pi a^2 \sigma T^4.
\label{equilium}
\end{equation}
The calculated \lbol\ is $1.1\times10^{45}$~\lum\ given a dust distance $R=52$ light days
by the time lag to the optical peak. 
The value is slightly higher yet comparable to the observed peak \lbol\ (\lpeak) 
of $7\times10^{44}$~\lum, providing valid evidence for the dust echo interpretation.

\subsection{Evaporation of Torus Dust and the \feii\ Emergence}

The torus inner radius can be computed by the dust sublimation radius (\rsub) below 
(here we used the formula given in \citealt{NU2016}):
\begin{eqnarray}
\rsub & = &  0.121~pc \left(\frac{\lbol}{10^{45}\;\mathrm{erg\;s^{-1}}}\right)^{0.5}\left(\frac{\tsub}{1800\;\mathrm{K}}\right)^{-2.804} \left(\frac{a}{0.1\;\micron}\right)^{-0.510}, \label{eq:Rsubiso}
\end{eqnarray}

The original torus inner radius is determined by the \lbol\ before the primary outburst,
whose corresponding \rsub\ is $5.6\times10^{-2}$~pc, that is 66 light days.
As the luminosity of PS1-10adi increases, the illuminated dust in the inner region
evaporates gradually, pushing the inner side of dusty torus outward.
The calculation at the optical peak yields a new \rsub\ of 0.10~pc
($\sim120$ light days), so the pre-existing dust located between 66 and 120 light days 
should have been evaporated. 

The dust sublimation as a function of temperature can be estimated by

\begin{equation}
  \frac{dm}{dt}=-\rho~A~\nu_{0}~\left (\frac{\mu}{\rho}\right )^{1/3}~e^{-B/kT}
\label{subrate}
\end{equation}

We adopt $\rho=2.5~\rm g~cm^{-3}$, $\nu_{0}\simeq10^{15}~\rm s^{-1}$, $B/k=7\times10^4$~K, 
$\rho/\mu=10^{23}~\rm cm^{-3}$ following Equation 7 of \citealt{Lu2016}, which are
representative values for refractory grains (\citealt{GD1989}; \citealt{WD2000}).
The A is the total illuminated dust area, that is $1.7\times10^{35}~\rm cm^2$ at E2
($A=L_{\rm dust}/\sigma T^4$), giving a dust sublimation rate 
$dm/dt=1.0\times10^{26}~\rm g~s^{-1}$.
The precise mass of evaporated dust is difficult to acquire because of the poor sampling
of the light curves.  We hypothesize that the sublimation rate at E2 is an average
value between -98 and +120 days (regions between the two orange dashed lines in 
Figure~\ref{bol}), the integrated evaporated dust is thus 
$\sim1.9\times10^{33}$~g, that is around 1~\msun. 

\begin{figure}
\centering
\includegraphics[width=8.5cm]{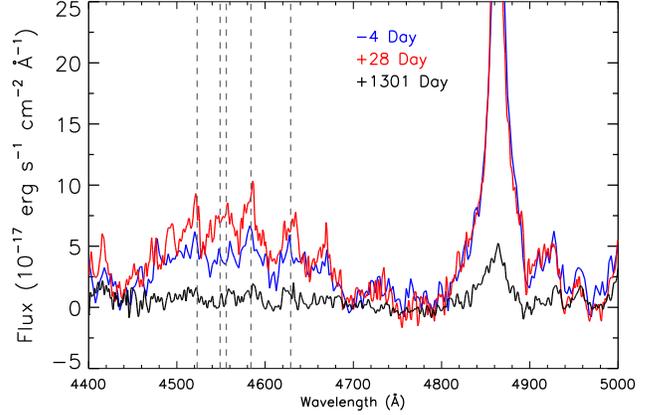}
\caption{
The variability of optical \feii\ emission of PS1-10adi (data drawn from K17).
Three spectra taken at -4 (blue), +28 (red) and +1301 days (black) 
respective to the optical peak are presented after the subtraction of continuum.
The position of \feii\ lines are marked as dashed grey lines.
The \feii\ intensity has strengthened at early stage (from -4 to 28 days) 
while faded over when the event ends (+1301 days).
}
\label{feii}
\end{figure}

Similar to the story told in PS16dtm (\citealt{Jiang17}), metals originally reserved 
in the evaporated dust will enter into the gas and may be ionized, giving rise to metal lines, 
such as \feii\ multiplets observed in PS1-10adi
(Figure~\ref{feii}, see also Supplementary Figure 2 in K17),
making it resemble a classical narrow-line Seyfert~1 galaxy.
If we believe that metals are mostly locked in the dust and the gas holds 
the solar metallicity, the gas mass should be higher than the dust mass by a factor 
of $\alpha\sim50$.
Supposing that the gas is distributed in the spherical region (outer radius 
$R_{\rm o}=120$~light days) with a covering factor of $f_{\rm c}$, the column density 
of the new broad-line region (BLR) can be calculated as below.

\begin{equation}
  N_{\rm H}=\frac{\alpha~m_{\rm dust}}{4\pi~f_{\rm c}~R_{o}^2~m_{\rm H}} 
\label{nhcal}
\end{equation}

The derived \colnh\ is $1.2\times10^{23}~\rm cm^{-2}$, in which we have adopted 
the $f_{\rm c}=0.4$ (see \S\ref{cf}). 
Both the gas mass and \colnh\ of new BLR here are comparable with 
typical AGNs (\citealt{Peterson1997}; \citealt{Netzer2013}),
indicating that they are likely sufficient to produce the observed \feii\ strength
even if the physical mechanisms are still not very clear to date.
It is worthwhile to note that the \feii\ multiplets have almost vanished after the 
outburst (see its +1301 days spectrum in Figure~\ref{feii}), agreeing well with
the photoionization model.

\smallskip
\smallskip

Many previous studies show that the photoionization can not fully address
the \feii\ emission of AGNs(\citealt{Collin2000}; \citealt{Sigut2003}; \citealt{Baldwin2004}) 
yet collisional excitation might also act (e.g., \citealt{Joly1981}; \citealt{Veron2006}).
This could be feasible in TDEs because the unbound debris ejected from the disrupted star 
may cause shocks by interacting with the gas and excite the iron. 
One needs to illustrate whether the time of collision is early enough to result in the 
instant presence of \feii.  
The maximum escaping velocity of the unbound debris 
$v_{\rm max}=(2GM_{BH}R_{\ast}/R^{2}_{t} )^{1/2}=9\times10^{3}~\rm km~s^{-1}$
in the case of one solar mass star disrupted by a $\sim 10^{7} M_{\odot}$ BH,
so the timescale of reaching the torus is $r_{\rm torus}/v_{\rm max}=6$~yr 
given a $r_{\rm torus}$ of 66 light days.
For the bound debris, although the circularization process is still not fully understood yet, 
several simulations show that the circularization timescale is in the range of several times 
to $\sim10$ times the orbital period of the most bound debris 
(\citealt{Bonnerot2016}; \citealt{Hayasaki2016}). 
The period of the most bound debris can be estimated by 
$T_{\rm min}=2\pi G\mbh~(-2 \epsilon_{\rm min})^{-3/2}$
in which $\epsilon_{\rm min}=-G\mbh~R_{*}/R^2_{t}$ is the lowest specific mechanical energy. 
The derived  $T_{\rm min}\sim0.6$ yr, meaning that the circularization timescale could be also 
as long as years yet with large uncertainties. 

In brief, we surmise that the rapid emergence of prominent \feii\ lines could be a universal
characteristic of TDEs in AGNs, which will potentially promote us to understand the 
origin of \feii\ emission in AGNs.

\section{Late-Stage Rebrightening}

\subsection{Late Accretion?}

\begin{figure}
\centering
\includegraphics[width=8.5cm]{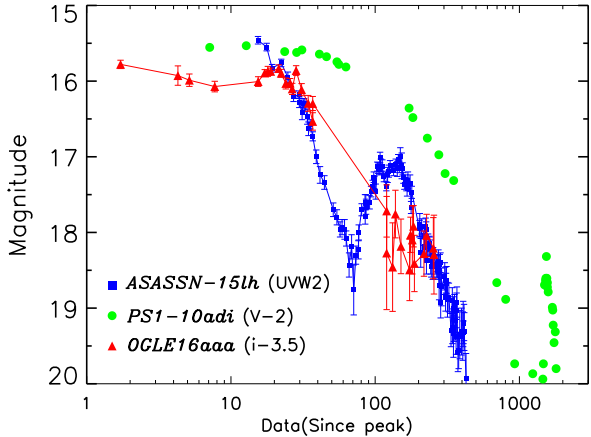}
\caption{
The comparison of the three TDE candidates with rebrightening-like light curves.
Apart from the late-stage rebrightening of PS1-10adi (green dots), 
ASASSN-15lh presents a giant double-peaked light curve
(blue squares, data from \citealt{Godoy-Rivera2017}) yet OGLE16aaa (red triangles, 
data from OGLE-IV Transient Detection System~(http://ogle.astrouw.edu.pl/ogle4/transients/2017a/transients.html) shows a slight rebrightening at earlier stage.}
\label{reb}
\end{figure}

The rebrightening hump appeared in the UV-optical light curve is non-negligible 
while it has not been discussed in K17.
As it happens, the similar hump is also present in the mid-IR light curve, with
W1 band ahead of W2. 
Besides that, the X-ray emission has been first detected with a $0.3-10$~keV luminosity
of $2\times10^{43}$~\lum during the rebrightening (see Supplementary Table 4 of K17).
Unfortunately, the photons are too few to constrain whether it's dominated
by soft X-rays or not. 

First, we attempt to naively understand the rebrightening as a late-stage accretion.
The peak \lbol\ of the rebrightening is lower than the primary outburst
by one order of magnitude, the time delay of dust emission relative to
the rebrightening should be $\sim100$ days as long as no dust newly formed.
The fitted \tdust\ at E6 is $1.7\times10^{3}$~K and then it has dropped quickly
to $\sim6\times10^2$~K at E7. The mid-IR emission at E6 requires
\lbol$\approx1.7\times10^{45}$~\lum\ according to equation~\ref{equilium} , that is
two orders of magnitude higher than the observed \lbol\ at $\sim100$ days before E6.
One way may resolve the conflict is that the blackbody assumption for 
dust emission could be over-simplistic and the dust absorption efficiency 
($Q_{\rm abs}\propto \nu^{\beta}$) should be taken into account.
The $\beta$ value is quite uncertain: it approaches zero for very large grain (grey case);
while for small grains, $\beta$ strongly depends on the composition, i.e.
$\simeq2$ for graphite, $\simeq1$ for silicate, and $\simeq-0.5$ for SiC.
If we adopt $\beta=2$ (see formula 8.16 in Kruegel 2003):
\begin{equation}
  \frac{L_{\rm bol}}{16\pi^2 R^2} \simeq 1.47\times10^{-6}aT^6,
\end{equation}
we get $\tdust=1.0\times10^3~K$, calling for \lbol\ of
$1.3\times10^{43}$~\lum\ for dust with MRN size distribution (\citealt{Mathis1977}),
that is still comparable with the observation. However, it is hard to imagine that 
the dust properties responsible for E2 and E6 are so different. The other negative 
evidence stems from the concomitant X-ray emission, which is strangely absent in the
early accretion but not in the late accretion. Therefore, we conclude that 
the late-stage accretion scheme is very unlikely.

\subsection{Rebrightening in Other TDEs}
\label{rebright}
We have noticed that among reported TDEs, ASASSN-15lh and OGLE16aaa 
also show a rebrightening characteristic in their UV-optical light curves (see Figure~\ref{reb}). 
In fact, it's exactly the giant double-humped feature that motivated reconsideration
of the nature of ASASSN-15lh, which was first claimed to be an unprecedentedly
luminous SNe in view of its spectral resemblance to 
hydrogen-poor super-luminous SNe (SLSNe) at early stage (\citealt{Dong2016Sci}).
Together with several other lines of evidence by further monitoring
(e.g., temperature evolution), ASASSN-15lh can be better understood as a TDE
occurred in a spinning BH with mass of $3\times10^{8}$~\msun\
(\citealt{Leloudas2016}; \citealt{Kruhler2018}) rather than a SNe.
Coincidentally, the X-ray emission hasn't been detected in ASASSN-15lh either until 
the rebrightening, which is about 4 months after the primary peak (\citealt{Margutti2017}).
Both the circulization (\citealt{Guillochon2015}) and 
reprocessing (\citealt{Metzger2016}) models have been suggested possible to 
address the intriguing UV-optical light curve (\citealt{Leloudas2016}).
The rebrightening of OGLE16aaa is visible, yet much less obvious as ASASSN-15lh.
\citealt{Wyrz2017} has tried to explain the variability as
induced by a binary BH on a tight orbit or due to disc precession 
or circularization on a timescale of about a month.
\emph{Swift} observations taken around the same time of optical flare
detected no X-ray emission. However, further \emph{Swift} exposures at $\sim 140$ 
dates later enabled us to detect significant X-ray emission from this
event, which then decayed and faded over a few months (\citealt{Auchettl2017}).

Both the rebrightening as well as associated X-ray onset 
in ASASSN-15lh and OGLE16aaa are similar to PS1-10adi, indicating plausibly 
a common and unified phenomenon in TDEs waiting to be understood.
However, PS1-10adi is still somehow different from the other two.
First, its occurring time of the rebrightening relative to the primary outburst
is much later (see Figure~\ref{reb}), at some 1500 days when the TDE emission has almost 
disappeared.
Second, the pc-scale gas and dust around the SMBH of PS1-10adi should be pretty ample, 
evidenced by the intense usual AGN activity and high dust covering factor 
revealed by the dust echo (see S\ref{energy}).
In contrast, the nucleus of ASASSN-15lh and OGLE16aaa are relatively quiescent despite
of weak AGN signatures (\citealt{Kruhler2018}; \citealt{Wyrz2017}).

\subsection{Interaction between Debris or Outflow and Torus}
\label{outflow}

\begin{figure}
\centering
\includegraphics[width=9.5cm]{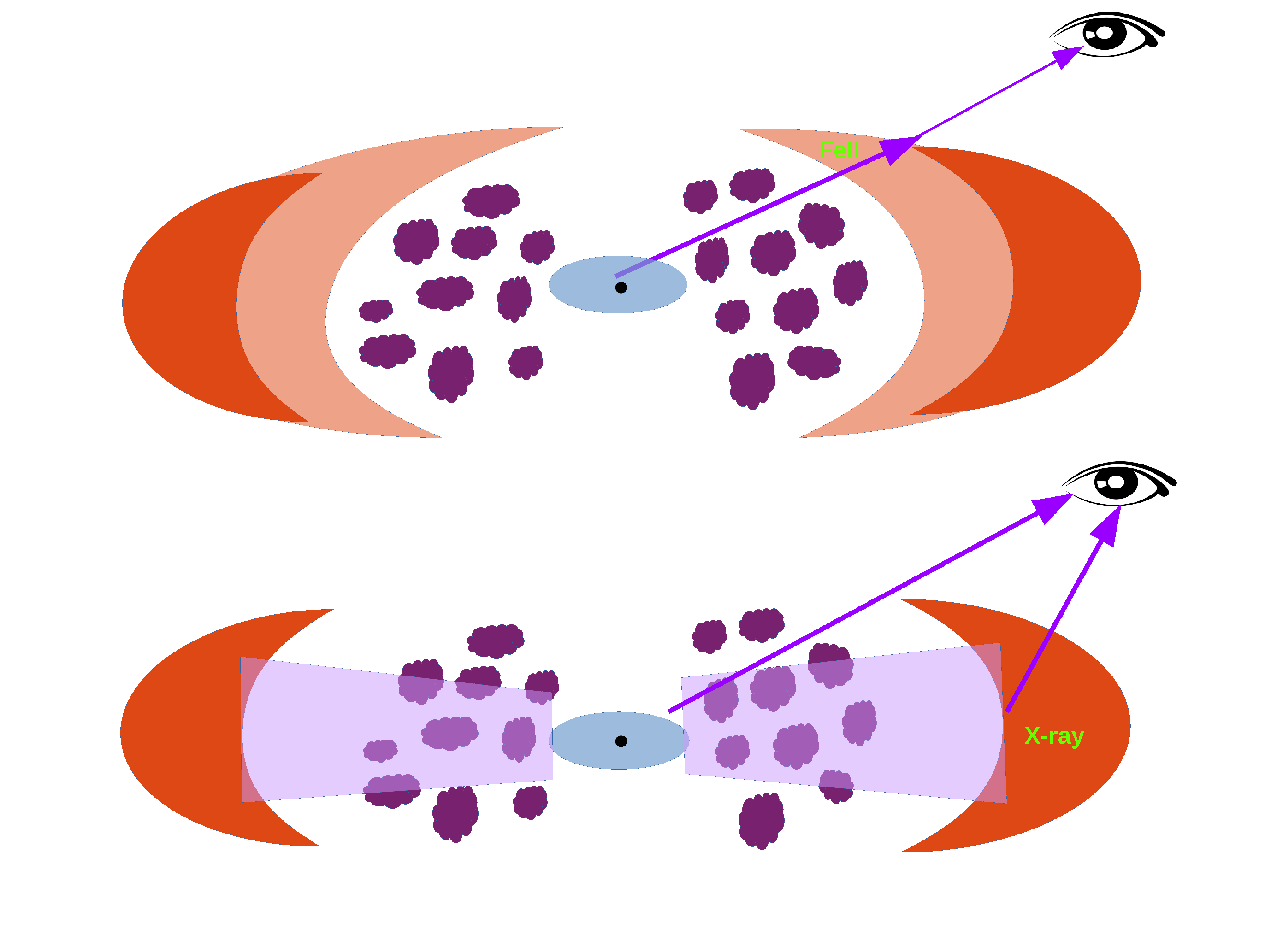}
\caption{
Schematic of the PS1-10adi model.
In the standard AGN unification, SMBH (black solid circle), 
accretion disk (blue ellipse), BLR (aubergine clouds) and torus (orange tori)
are distributed from inside to outside.
Top panel: the central engine of PS1-10adi was partly obscured by torus
yet largely exposed after the TDE due to the receding of torus.  The metals 
released from the evaporated dust (light orange region) will give rise 
to the observed strong \feii\ lines.  
Bottom panel: the high-speed outflow (denoted as light purple) launched 
from the accretion disk collides 
with the torus and produce the late-stage rebrightening and X-ray onset.
}
\label{model}
\end{figure}
\smallskip

The particularities of PS1-10adi encouraged us to come up with possible new mechanisms
of the rebrightening.
Could it be caused by the collision between the unbound stellar debris 
and the interstellar medium (ISM) at scale of AGN torus?
The spread of specific orbital energy $\Delta \epsilon=GM_{\rm BH}R_{\ast}/R^{2}_t $ 
is almost a constant for a  large range of 
$\beta \equiv r_{t}/r_{p}$ (\citealt{Stone2013}; \citealt{Tejeda2017}). 
The total orbital energy of unbound debris is 
$\Delta E < 0.5 M_{\odot} \times 0.5 \Delta \epsilon$. 
The interaction timescale for this collision is $t_{\rm intact} \sim 0.5 r_{\rm torus}/0.5v_{\rm max} = r_{\rm torus}/v_{\rm max} \sim 1500$ days. 
Therefore we can estimate the upper limit of rebrightening power assuming that all of 
the orbital energy can be transferred into the gas internal energy and radiate efficiently: 
$L_{X}({\rm max}) < \Delta E/t_{\rm intact}=1.5\times 10^{42} ~\rm ergs~s^{-1}$, 
which is one order of magnitude lower than observations. 
As a result, we argue that the rebrightening may not be due to the collision 
between the unbound debris and the torus.

Apart from the ejected debris, outflows launched by the TDE 
can also serve as a source of collision (see Figure~\ref{model} as an illustration).
If that is the case, the observed outflow velocity can be estimated by
$v_{\rm ej}=r_{\rm torus}/t_{\rm reb}\sim120/1500*c$=$2.4\times10^4$~\kms.
Outflows with such a kind of high velocity has been found in observations
(e.g., \citealt{Blanchard2017}; \citealt{Kara2018}), 
and numerical simulations in which the outflow may be produced during the self-crossing 
process under the general relativistic precessing  
(e.g., \citealt{Sad2016}; \citealt{JiangYF2016}), or from the final 
accretion disk with super-Eddingtion accretion rate (e.g., \citealt{Dai2018}). 
For example, in the simulation of \citealt{Sad2016}, about 10\% of the star's total mass 
can be transferred into outflow after the violent self-crossing, 
with outflowing velocity of $\sim$ 0.1c, (yet we should stress that the simulated 
objects therein is 0.1 solar mass star tidal disrupted by a BH of $10^5$~\msun). 
Hence the total kinetic energy of this outflow is in the order of $10^{50}$ ergs. 
As a comparison, the integrated UV- optical bolometric light curve between 
1500 and 2000 rest-frame days results in a total energy of $\sim5\times10^{50}$~erg, 
which is on the same order of magnitude with the self-crossing outflow. 
In the torus interacting model, both the primary collisional and dust reprocessed emission 
are almost local, which can naturally address the high \tdust\ at the rebrightening stage.
This is also supported by the time lag between the optical and IR 
rebrightening, which is shorter than the light travel time from BH to 
the new torus boundary ($\sim100$ days).

Last, we caution that besides the torus dust distributed in the equatorial plane around 
the accretion disk, dust elongated in the polar direction has also been found in both 
type 1 and type 2 AGNs via recent mid-IR interfermetry observations 
(\citealt{Honig13}) and SED fitting (\citealt{Lyu2018}).
It's suggested that a radiatively driven dusty wind, possibly launched in a puffed-up region
of the inner hot part of the torus, is responsible for the polar dust (\citealt{Honig12}).
The dusty wind is mostly optically thin and extends in the polar direction
over parsecs or even tens of parsecs.
On the other hand, the polar outflow is not inhibited inherently and has been tentatively
reported in some broad absorption line quasars (e.g., \citealt{Zhou06}).
Therefore, the collision between the polar outflow and the polar dust can not 
be excluded completely to produce the observed rebrightening.

\section{Revisiting the Nature of PS1-10adi and Analogs}
\label{energy}

In previous sections, we have assumed that the nuclear transient in PS1-10adi is a TDE
and have tried to understand all observations in the context of TDEs. 
However, as mentioned in K17, the case of extraordinary bright SNe occasionally 
exploded in the galactic nucleus is still impossible to disentangle from the TDE 
case up until then. In this section we will try to revisit the nature of PS1-10adi 
and other similar events with aid of mid-IR light curves.

\subsection{Mid-IR Light Curves of PS1-10adi Analogs}

\begin{figure}
\centering
\includegraphics[width=8.5cm]{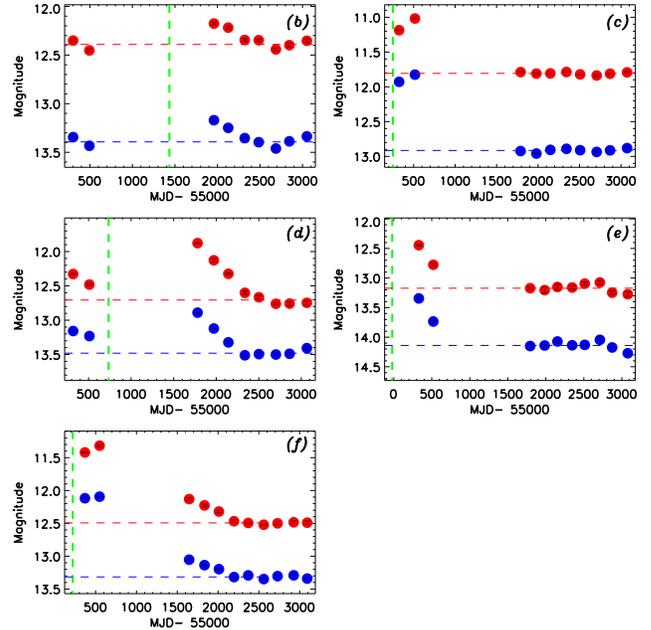}
\caption{The \wise\ W1 (blue) and W2 (red) light curves of PS1-10adi analogs reported in K17.
b-f: PS1-13jw, CSS100217, J094608, J094806 and J233454.
The green dashed line marks the optical peak date.
}
\label{k17lc}
\end{figure}

Besides PS1-10adi, K17 has reported 5 other outbursts in the centers of active galaxies. 
They constitute the sample of highly-energetic nuclear transient events along with PS1-10adi. 
To understand their nature as a whole, we have retrieved their WISE and NEOWISE photometric 
data and presented the light curves in Figure~\ref{k17lc}. Significant mid-IR flare-like
variability is visible for them all. As we already know the optical peak date of these events, 
we first estimated the background emission unassociated with the outbursts by averaging
the magnitudes before or after the flares when the light curves maintain a stable level.

It's interesting to note that a sustainedMIR excess is present in J094608 before 
the optical flare. To exclude the probability of spurious signal caused by a systematical bias 
between ALLWISE and NEOWISE-R data, we have performed aperture photometry on the 
time-resolved coadds of \wise\ images (\citealt{Meisner2018}). 
The excess ahead of the optical flare is proven to be real. 
It may be simply attributed to a higher dust covering factor then as the torus has
receded to a further distance after the outburst, but it may be also reconciled 
by several alternative modes under the TDE scenario. \citealt{Guillochon2015} suggested
that there is a delay between the flare and the disruption of the star of order years 
for BH mass of $10^7$~\msun. If the unbound debris collides with (torus) ISM, producing
shocks and emitting UV/X-rays, it may lead to the enhanced IR emission before the primary outburst. 
This may be also a potential mechanism to address the early IR brightenings 
in PS16dtm (\citealt{Jiang17}) and PS1-10adi (see \S\ref{echo}). 
Other explanations, such as invoking stream collisions as a precursor to the 
main TDE flare (e.g., Bonnerot \& Rossi 2018)  or double TDEs induced by stellar binaries
incident on SMBH binaries (Coughlin et al. 2018), are also possible. 
Owing to the deficiency of observational data before the event (e.g., X-rays),
a thorough discussion of these possibilities is beyond of this work 
but could be tested in the future for TDEs with multiwavelength monitoring.

\subsection{Infrared Luminosity Compared with Known Supernovas}

The SNe scenario has been discussed by K17 in parallel with the TDE case.
Spectroscopically, PS1-10adi bears similarity to type~IIn SNe 
(SNe IIn, \citealt{Schlegel1990}; \citealt{Fili1997}),
which is distinguished by relatively narrow emission lines and slowly declining light curves.
The signatures of SNe IIn are not associated with the explosion itself, but rather
with the interaction between ejecta and dense circumstellar medium (CSM) generally
produced by pre-SN mass loss.
SNe IIn have gained considerable attention over the past decade partly because they show 
a huge range in bolometric luminosity (e.g., \citealt{Richardson2014}; \citealt{Li2011}),
among which the SLSNe are of greatest interest
(e.g., \citealt{Smith2007}; \citealt{Ofek2007}; \citealt{Gal2012}).
It's firstly necessary to demonstrate why PS1-10adi and analogs prefer to
be linked to AGNs if they are indeed SLSNe.
One possible formation channel of these SLSNe has been proposed to be the runaway mergers 
of massive stars in dense and young stellar clusters, giving rise to very massive
H-rich CSM (\citealt{Por2007}),
which seems feasible in high-density central regions of an active galaxy.
The dense AGN environment can also provide suitable conditions 
for high ISM pressure as well as the photoionizing radiation to 
trap a large fraction of the mass lost from the progenitor (\citealt{Mackey2014}).

\begin{figure}
\centering
\includegraphics[width=8.5cm]{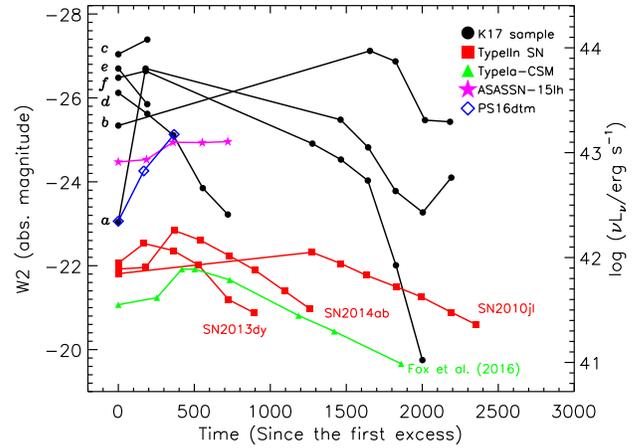}
\caption{
The \wise\ W2 (4.6$\mu$m) absolute magnitude (background-subtracted) of K17 sample
and SNe with prominent mid-IR excesses. a-f: PS1-10adi, PS1-13jw, CSS100217, J094806,
J094608 and J233454, namely the six objects listed in the supplementary table 1 of K17.
The Type-IIn SNe are plotted in red squares and Type Ia-CSN SNe is shown as
green triangles (\citealt{Fox2016}).  
We have also overplotted ASASSN-15lh (magenta star) and PS16dtm (blue diamonds) 
for comparison.
}
\label{lum}
\end{figure}

Could the SNe scenario also meet the mid-IR flare freshly 
revealed by the \wise\ light curve?
\citealt{Fox2011} have detected 10 out of 68 SNe IIn with late-time (>100 days)
IR emission from a warm \emph{Spitzer}/IRAC survey, holding a maximal IR luminosity
of $10^8\sim10^9$~\lsun, that is around $10^{42}$~\lum.
Statistically, the IR emission of SNe IIn are somewhat more luminous and 
more sustained than other types (e.g., type~Ia), due to the heating of 
pre-existing dust by radiative shocks between the expanding SN shell 
and the dense wind of the progenitor (\citealt{Fox2013}; \citealt{Tiny2016}).
Even so, the peak IR (W2 band) luminosity (\lw) of PS1-10adi and analogous objects are 
uniformly higher than SNe~IIn by at least one order of magnitude, exceeding $10^{43}$~\lum.
As a comparison, the \lw\ maximum of the other AGN transient PS16dtm up to 2017 July
(latest public epoch) is also above $10^{43}$~\lum\ (\citealt{Jiang17}) already.
We caution that previous \emph{Spitzer} follow-ups (\citealt{Fox2011}) may missed the
early IR emission, so then we have subsequently checked the \wise\ light curves of all SNe~IIn
exploded during 2009 and 2016 from the open supernova catalog~\footnote{https://sne.space/}
(\citealt{Guillochon2017}). 
Among them, the most IR luminous ones (e.g., SN2010jl, SN2013dy, SN2014ab) show no
higher luminosity and still possess \lw\ $\sim10^{42}$~\lum\ (see Figure~\ref{lum}).
As we mentioned in \S\ref{rebright}, the controversial event ASASSN-15lh was discovered 
as a historical SLSNe at first while more observational facts later favour the TDE scenario. 
A mid-IR flare after the giant optical outburst has also been detected 
in ASASSN-15lh by \wise\ with peak \lw\ $\sim10^{43}$~\lum,
making it stand out from SNe despite slightly lower than PS1-10adi.
In one word, the IR luminosity of PS1-10adi and analogs are significantly 
higher than known certified SNe IIn as well as any other types of SNe to our knowledge.

\subsection{Dust Covering Factor: Consistent with AGN Torus}
\label{cf}

The extremely high IR luminosity of PS1-10adi can not reconcile with SNe, 
but seems fairly practical for TDEs in AGNs, in which the dusty torus around the BH will
unavoidably absorb the TDE emission and reprocess them into the IR.
To depict the dust content around PS1-10adi more quantitatively, 
we try to calculate its dust covering factor (\fdust), 
namely the fraction of the sky obscured by dust when seen from the central source.
The idea is very straightforward that the original UV-optical photons
towards the dust region will be entirely absorbed and re-emitted in the IR.
Therefore, the ratio between energy radiated through dust reprocessing (\edust) and
total energy initially released by the event (\erad) can be used to indicate \fdust\
(see also \citealt{vanv2016}).
K17 has obtained a \erad\ of $\sim2.3\times10^{52}$~erg by integrating the 
UV-optical blackbody emission and taking into consideration of the early UV excess 
and missing rise-time contribution.
Unfortunately, the precise calculation of \edust\ is hindered by the sparse sampling 
and the 3-year-gap of the mid-IR light curve.
As an acceptable alternative, we have calculated the \edust\ basing on the 
piecewise linear luminosity evolution, yielding a value of $1.0\times10^{52}$~erg.
The derived \fdust\ of PS1-10adi is thus $\sim0.4$, that is 
broadly consistent with AGNs estimated by other methods,
such as fitting torus component to the SEDs
of AGNs (e.g., \citealt{Fritz2006}; \citealt{Mor2009}; \citealt{Roseboom2013}).
We have subsequently computed the \fdust\ for the left 5 events in K17 using
the same approach and obtained comparable values.

The high \fdust\ of PS1-10adi is a natural consequence of AGNs, which is automatically 
satisfied in the TDE case as it happens in the accretion disk scale encircled by the torus.  
Core-collapse SNe have been believed to be one of the major production sites
of cosmic dust grains (see a recent review by \citealt{Sarangi2018}), 
either by ejecting metal-rich materials into space for dust condensation or 
through the interaction between the ejecta and dense CSM. 
However, the amount of dust produced by an individual SNe is very limited and hard
to manifest the SNe as a such high covering factor.
The only conceivable SNe likelihood remained to be a SLSNe exploded inner or exactly
in the torus (e.g., \citealt{Assef2018}). In the meantime, the torus or AGN gas may 
also provide the dense CSM that is thought to be needed to make the SNe super-luminous. 
The SNe scheme can not be ruled out completely with the caveat that the SNe explosion
within the torus scale hasn't been well demonstrated.

We do not expect that the mid-IR echoes have the function to definitely diagnose 
the nature of PS1-10adi and analogs, but they indeed confine the outburst location 
to be no larger than the torus scale confidently.
It's not trivial to differentiate TDEs and SNe with existing information and our current
knowledge. Some other studies have even invoked peculiar AGN processes to generate
these intense nuclear flares of Seyfert galaxies. 
For instance, the interaction between BH disk winds and BLR clouds 
is proposed to reproduce the observed luminosity and timescale (\citealt{Moriya2017}).
In fact, a high-amplitude AGN variability could be misidentified as TDE or vice verse
(e.g., \citealt{Merloni2015}; \citealt{Saxton2018}).
In this way, the ionizing continuum and mid-IR echoes may be the least informative aspect
of the transient, and other identifying signatures (e.g., chemical abundance ratio,
\citealt{Yang2017}) are needed.

\section{Summary}

The investigation of TDEs in AGNs has become more and more realizable thanks to the 
rapid development of time-domain astronomy nowadays.
Continuing with our first dust echo case study on PS16dtm (\citealt{Jiang17}),
which is a TDE occurring in a Seyfert~1 host found by Pan-STARRS, 
we have next turned our attention to the other event PS1-10adi (K17) discovered 
by the same survey.
The outburst of PS1-10adi happens much earlier than PS16dtm and thus yields a more complete 
light curve, that will provide us an excellent opportunity to understand TDEs in AGNs.

The mid-IR echoes appeared in both events display some distinctive characteristics by comparing
with previous TDEs in quiescent galaxies. 
The complete understanding of these two events is beyond this work, yet their 
unusual behaviors can be largely attributed to a pre-existing dusty torus in AGNs but 
absent in quiescent galaxies.
First, the mid-IR flares are detectable even earlier than the optical ones.
The dust temperature naively fitted by blackbody emission in the early stage is 
higher than the dust sublimation temperature, indicating strongly on-going dust evaporation.
We infer that it's the torus dust around the BH that have (partially) obscured and weakened 
the optical emission, which has been conversely transferred to the IR emission.
Along with the rising of the TDE bolometric luminosity, the inner radius of the torus
receded until the peak luminosity. During this process, the metals originally contained 
in the dust will be delivered to the gas and give rise to the prominent \feii\ multiplets
seen in PS16dtm and PS1-10adi. 
The \feii\ lines of PS1-10adi have faded over when the outburst ends, 
supporting strongly the photoionizing mechanism, although 
the collisional excitation may also act, triggered by the interaction 
between the unbound stellar debris with the torus gas in the TDE scenario.

The other intriguing feature of PS1-10adi is the late-stage UV-optical rebrightening 
at $\sim1500$ rest-frame days after the primary outburst.  The peak luminosity of the hump is 
$\sim5\times10^{43}$~\lum, that is two orders of magnitude lower than the primary peak.
In agreement with most other optical TDEs, the X-rays hasn't been detected 
in the early stage (\citealt{Auchettl2017}). Interestingly, the \swift\ observation 
triggered during the rebrightening has captured the X-ray emission.
In addition, there also exists a mid-IR rebrightening immediately after the the optical one.
The late-stage accretion possibility is disfavoured by the ultra-high \tdust\ 
and X-ray emerging.
We have also noticed that at least two other reported TDEs (ASASSN-15lh and 
OGLE16aaa) show somewhat similar behavior yet the rebrightening takes place at much 
earlier stage. Unlike the reprocessing or circularization model suggested in ASASSN-15lh,
we have proposed a new mechanism responsible for the rebrightening inspired by
the fact of AGN phase before PS1-10adi occurs.
In this scenario, the collision between the high-speed outflow and the torus 
has resulted in the UV-optical rebrightening and X-ray onset.
We suspect that it may be a universal phenomenon for the TDEs in AGNs. 
Further follow-ups of PS16dtm and more other events can help us confirm the idea or not.

The mid-IR light curves analyzed in this work can also help us understand
the nature of these highly energetic transient events in the centers of active galaxies.
The peak IR luminosities of PS1-10adi and analogs are more than 10 times higher 
than all known IR-luminous SNe, but could be rare SNe exploded in the torus.
Moreover, their IR luminosity is also much higher than normal TDEs in 
quiescent galaxies (e.g., \citealt{Jiang16}; \citealt{vanv2016}). 
It can be easily understood as the dusty torus effect, which solely acts in AGNs.
The torus has obscured the central engine of PS1-10adi with a covering factor of $\sim0.4$
and radiatively transferred a large fraction of high energy photons to the IR band.

Tori is an essential ingredient in the AGN unification model to explain various
observational characteristics exhibited in different types of AGNs 
although its attributes are still far away from clarity.
TDEs in AGNs may enable us to probe the torus from a novel perspective.
The dust echo of the torus will respond to the tidal disruption radiation as an
outstanding mid-IR flare, which is presumably even remarkable than the optical flare.
The time delay of the flare in the mid-IR to the optical is an expression of the
torus physical scale and the total radiated energy in the IR can be used to
measure the dust covering factor of AGN.
Similar phenomenon has also been reported in the other sorts of systems undergoing
dramatic change of accretion rate, namely changing-look AGNs in which
the AGN type has transited from type 1 to type 2 or vice versa (\citealt{Sheng2017}; 
2018 in preparation).
Moreover, the interaction between the high-speed outflow and ISM may also
give useful insights into the torus information.
The new approach demands timely trigger of mid-IR monitoring (e.g., \spitzer) 
of the TDE candidates found in AGNs by X-ray or optical band. On the other hand, the
mid-IR flares identified from surveys like \wise\ (Jiang et al. in preparation) also 
deserve prompt follow-ups in other bands (e.g., X-rays). The complete coverage of all
events at each band is too expensive and unrealistic, yet could be achieved by 
certain well-defined projects at the upcoming golden era of time-domain astronomy.
\\

\acknowledgements

We thank for the anonymous referee for many constructive comments and suggestions.
We thank  Dr. Erkki Kankare for providing us the optical spectra data of PS1-10adi.
This work is supported by the National Basic Research Program of China
(grant No. 2015CB857005), NSFC (NSFC-116203021, NSFC-11703022, NSFC-11833007), 
Joint Research Fund in Astronomy (U1431229, U1731104) under cooperative agreement
between the NSFC and the CAS, Anhui Provincial Natural Science Foundation and 
the Fundamental Research Funds for the Central Universities.
This research makes use of data products from the
{\emph{Wide-field Infrared Survey Explorer}, which is a joint
project of the University of California, Los Angeles,
and the Jet Propulsion Laboratory/California Institute
of Technology, funded by the National Aeronautics and
Space Administration. This research also makes use of
data products from {\emph{NEOWISE-R}}, which is a project of the
Jet Propulsion Laboratory/California Institute of Technology,
funded by the Planetary Science Division of the
National Aeronautics and Space Administration.

\end{document}